%% file: master.tex
\documentclass{llncs}
\pdfoutput=1

\usepackage{epsfig,endnotes}
\usepackage{setspace}
\usepackage{amssymb}
\usepackage{amsmath}
\usepackage{graphicx}
\usepackage{url}
\usepackage{array}
\usepackage{subfigure}
\usepackage{multicol}
\usepackage{fancyvrb}
\usepackage{multirow}
\usepackage{xspace}
\usepackage{color}
\usepackage{xcolor}
\usepackage{dsfont}

\newcommand\equalhat{%
\let\savearraystretch\arraystretch
\renewcommand\arraystretch{0.3}
\begin{array}{c}
\stretchto{
    \scalerel*[\widthof{=}]{\wedge}
    {\rule{1ex}{3ex}}%
}{0.5ex}\\ 
=%
\end{array}
\let\arraystretch\savearraystretch
}

\title{Detecting Behavioral and Structural Anomalies in MediaCloud Applications}

\author{Guido Schwenk \and Sebastian Bach}
\institute{	Technische Universit\"at Berlin, Marchstr. 23, 10587 Berlin}

\begin{document}

\maketitle


\input{sections/abstract.tex}
\input{sections/introduction.tex}

\input{sections/alumediacloud}

\input{sections/related_work}

\input{sections/learning_methods}

\input{sections/behavioral_anomaly_detection}
\input{sections/structural_anomaly_detection}

\input{sections/datasets}
\input{sections/evaluation}

\input{sections/conclusion}
\input{sections/acknowledgements}

\bibliographystyle{abbrv}

\bibliography{bib_sebastian,bib_guido}

\end{document}

%% file: sections/abstract.tex
\begin{abstract}
In the past years technological advances such as the increasing bandwidth in network infrastructures and new software developments such as message and agent-based systems gave rise to the field of cloud technologies, which have evolved from abstract concepts to concrete solutions, ranging from flexible, platform-independent systems to highly specialized software solutions. In this paper we introduce and evaluate two anomaly detection methods to achieve a higher level of security in a specific cloud solution for interactive media, the Media Cloud from Alcatel-Lucent \cite{CucObeSteDomMul13}. The Media Cloud focuses on real-time processing of interactive media applications, allowing for optimal resource planning using highly specific functional components. The proposed anomaly detection methods are designed to work complimentary to each other and are capable of detecting known and unknown vulnerabilities and security issues, offering very low false positive rates and very high detection rates, as is shown by the evaluation on real Media Cloud data and synthetic data. The proposed methods use behavioral and structural features, and are capable of locating the detected anomalies as well, giving the executing analyst easy insight into the running processes.
\end{abstract}

\keywords{Anomaly Detection, Cloud Security, Machine Learning, Graphs}

%% file: sections/introduction.tex
\section{Introduction}
\label{introduction}

In the last years, cloud services established themselves as a new and highly important technology. Due to the advantages of cloud services, noticeable in lower administrative costs and increased scalability, flexibility and availability, this growth will continue. There exist two main service categories of cloud computing which are relevant for this paper: the Platform-as-a-Service (PaaS) clouds and Software-as-a-Service clouds (SaaS). PaaS clouds are built of a large amount of virtual machines and provide access to middleware, run-time environments or complete operating systems. Customers of such services use those virtual machines to set up their own system, without having to care about maintenance and administration of an own server park. PaaS clouds also offer high flexibility. Virtual machines are easily set up with different operating systems, enabling an easier development, testing and service deployment for those platforms. Furthermore the number of managed systems can easily be scaled, enabling a flexible reaction to the demands of the market. In the pre-cloud era, such scalability was only possible with an own server park and not affordable to smaller or medium sized companies. Examples of PaaS clouds are the Amazon Elastic Cloud (\texttt{aws.amazon.com/de/ec2}) and the Google App Engine (\texttt{developers.google.com/appengine}). SaaS clouds have a higher abstraction level than PaaS clouds but offer the same kind of scalability. Instead of providing raw platforms to the customers, a SaaS cloud offers a selection of pre-installed software services. So the customer does not need to handle the installation and management of the operating system. Obviously this comes at the cost of lower flexibility, as only the functions of the pre-installed software services can be used. Examples of SaaS cloud services are the Amazon Web Services (\texttt{aws.amazon.com}) or dedicated media-streaming services like Watchever (\texttt{www.watchever.de}). More details on cloud technologies are given in the report of \cite{CSA13.1}. 

In this paper we develop and evaluate detection methods for the Alcatel-Lucent Media Cloud, a PaaS and SaaS cloud solution for media streaming. Its novel features, illustrated in Section \ref{alcatellucentmediacloud}, are not yet available on any other PaaS or SaaS cloud service, combining the flexibility and scalability of classical PaaS systems with the ease of use and high performance of dedicated media-streaming SaaS-solutions. While there exist approaches to secure common cloud technologies, those solutions can not be applied directly to the media cloud infrastructure. 
The contributions of this papers are the presentation of two complimentary methods for the detection and the localization of anomalies in the structure and in the communication behavior of Alcatel-Lucent Media Cloud applications. The method for the detection of behavioral anomalies uses features of the communication of individual components, while the method for the detection of structural anomalies is based on information of the application graphs. The evaluation of our method on a set of data of a real Media Cloud application shows very low false positive rates and true positive rates of up to $100\%$, while the evaluation on synthetic data shows reasonable results as well.
The rest of this paper is structured as follows: In Section \ref{alcatellucentmediacloud} we present the layout of the Alcatel-Lucent Media Cloud, its applications and its vulnerabilities. In Section \ref{relatedwork} the related work in the fields of cloud security and anomaly detection methods is discussed. After giving a short introduction to the utilized learning methods in Section \ref{sec:learningmethods}, the detection methods for behavioral and structural anomalies in Media Cloud applications are elaborated in Sections \ref{sec:behavioralanomalydetection} and \ref{sec:structuralanomalydetection}. The data sets and the evaluation are presented in Sections \ref{sec:datasets} and \ref{sec:evaluation}, followed by the concluding Section \ref{sec:conclusion}.

%% file: sections/alumediacloud.tex
\section{Alcatel Lucent Media Cloud}
\label{alcatellucentmediacloud}
Combining features of PaaS and SaaS cloud, the Media Cloud offers application providers a platform for easy development and deployment of large-scale internet services. Consumers are offered a large number of services, and application service aggregators and cloud service providers are offered an API and platform for easy composition of similar services. The main focus of Media Cloud applications is currently interactive media based applications like video conferences and video calls. Those services do not tolerate delay or lag in the streaming content. Optimizing the distribution of application components in reaction to dynamically varying user scenarios has therefore to fulfill very tight constraints. The following section explains those details of the layout and functionality of the Media Cloud which are relevant for security issues.

\subsection{Media Cloud Layout}

The Media Cloud is a framework with the objective to allow the execution of software services on a widely distributed network of computational resources. Its main focus is media streaming with a high level of interactivity and real-time applicability. Its distinctive feature
is the possibility to re-map the software components of an application very fast to different computational resources (i.e. servers), by using extremely specialized and minimized virtual machines, called components. As a result, the computational burden of each application can be moved exactly to computational resources which are closest to the respective consumer, or to those with
a higher capacity. This results in a very high quality of service, which is not only desirable, but necessary for most real-time media applications. An application is built by an interconnected set of those components which communicate with each other using a message system. Each application contains a core of components which realize the main function of this application. For a video chat application those might be components for recording, encoding, transferring, decoding and displaying video content. Since dynamic behavior during runtime is encouraged, however, additional components can easily be added. For the previous example components for filtering or capturing might be added, for example. Different application types can use the same type of component, but the way each application uses this component can differ during run-time and is defined by the purpose of its application. 

\subsection{Media Cloud Vulnerabilities}
\label{sec:mediacloudvulnerabilities}
Although the components and applications of the Media Cloud are developed under strict rules of quality assurance, there exist possibilities for different vulnerabilities, mostly caused by insecure application components. From the cloud computing vulnerabilities listed in the literature \cite{CSA13.2,VelZla11,MatKumLat09}, the following are relevant for the Media Cloud. 

\paragraph{Data breaches} pose a threat for cloud frameworks with shared resources. They are conducted by using side channel timing information to extract information from virtual machines on the same server, as described in \cite{ZhaJueReiRis12}, or via an insufficiently secure implemented multitenant cloud service database. The vulnerability for this threat is quite low for the Media Cloud, because generally its components do not share resources on servers. This could change in future versions, though. In a running Media Cloud application, data breaches could occur in form of an unusual communication behavior of single components in terms of message rate or the amount of traffic, used for transferring data from secured channels, or in form of unusual connections between individual components, used for example to transfer data via a specifically vulnerable component interface.

\paragraph{Account or Service Traffic Hijacking} via Phishing or Fraud is a risk for the Media Cloud. Although securing the provided credentials and account information is a responsibility of the application service provider or the consumer, one can not rule out a malicious exploitation of software vulnerabilities of Media Cloud applications. This correlates directly to insecure interfaces and APIs (e.g. authentication, access control, encryption). This could be used to maliciously access media streams, e.g. for eavesdropping on a confidential video conference. In the running media cloud application such malicious behavior could be visible as unusual behavior of a specific participant of the video conference, as he tries to access participants or application components previously not contacted.

\paragraph{Denial of service attacks} (DoS) on system resources, e.g. processing power, memory, disk space or network bandwidth, can decrease service quality or shut down the system completely. The sophisticated load balancer included in the Media Cloud takes care of the consequences, but does not find the origin of such attacks, as can be done by our proposed methods. In common network topologies a DoS attack is executed by using legal actions like HTTP-requests or addressing software interfaces, thereby accessing the assigned resources at full capacity to decrease its quality of service. If this attack is distributed between multiple attackers, the effect is even worse. In the Media Cloud framework multiple compromised resources, components or applications could be used for sending a malignant amount of messages to exhaust the application resources, rendering the attacked application or resource useless. Again this kind of attack is detectable in anomalous structures and anomalous patterns of behavior in the communication of the application.

%% file: sections/related_work.tex
\section{Related Work}
\label{relatedwork}

Before introducing our proposed anomaly detection method, we start with a discussion of some related work in the fields of component-based cloud computing and anomaly detection methods.

\subsection{Component-based Cloud Computing}
\label{componentbasedcloudcomputing}
The unique features of the Alcatel-Lucent Media Cloud, compared to other PaaS and SaaS solutions, are the systematic use of atomic components and the capability for real-time re-mapping component locations for an optimal distribution in terms of short transmission paths. To our best knowledge there exists no cloud solution handling the re-mapping in a comparable way. There exists, however, one solution which also makes use of atomic components, namely the 4caast project (\texttt{4caast.morfeo-project.eu}). Its scope is much wider than that of the media-centric Media Cloud, though, and there are no data sets available to test our methods on.

\subsection{Anomaly Detection Methods}
\label{anomalydetectionmethods}
While there exist methods for preventing some of those vulnerabilities relevant for the Media Cloud, e.g. guidances for secure clouds as in \cite{JanGra11,SulBonFurOrr13}, they are not capable of detecting unknown behavioral anomalies of Media Cloud components. A proven approach to this problem are anomaly detection methods, which have been developed in the field of intrusion detection. The anomaly detection paradigm relies on the possibility to learn a model only over normal data, i.e. data which represents normal behavior.  Once trained the learned model is capable of detecting behavior that deviates from the normal data and can therefore be classified as anomalous. Anomaly detection methods have been used for the detection of anomalies in network traffic \cite{BarKliPloRon02,LakCroDio05,LiCroDioGovIanLak06}, intrusion detection \cite{WanParSto06,IngIno07} or drive-by-download attacks \cite{CovKruVig10,SchBikKruRie12}, but also in a large amount of other areas, as nicely described by Chandola et.al. in \cite{ChaBanKum09}. 
Learning with kernels, as described e.g. by Sch\"olkopf and Smola \cite{SchSmo02} has been applied to different use cases. For our proposed structural anomaly detection method, specifically the latest research on graph kernels (e.g. by Hido et.al. \cite{HidKas09}) and research on the detection of anomalies in graphs (e.g. by Eberle et.al. \cite{EbeHol07}) are interesting. However, none of those approaches focuses on highly dynamic non-linear, non-planar graph structures, as is the case for Media Cloud applications.

%% file: sections/learning_methods.tex
\section{Learning Methods}
\label{sec:learningmethods}
Kernel methods like the (two-class) Support Vector Machine (e.g. \cite{cortes1995support,ShaChr04}) are known to perform very well in scenarios where both types of data (e.g. benign vs. malicious, including label information) are available and linearly separable. In the Media Cloud scenario, however, only the normally behaving data is available, collected during the normal usage of different applications. For this reason we decide to use unsupervised one-class support vector methods, which are shortly explained in the following sections. For further details on the selected methods, please refer to the cited references.

\subsection{One-Class Learning Methods}
\label{sec:oneclasslearningmethods}
To our best knowledge, there are two support vector kernel methods fit for one-class learning: The One-Class Support Vector Machine (OCSVM) \cite{scholkopf1999support} and a method called Support Vector Data Description (SVDD) \cite{tax2004support,changrevisit}. Both methods are unsupervised and aim at learning a model of normality which describes the training data they are presented with. The One-Class SVM is in its formulation very similar to the (two-class) Support Vector Machine, except it separates the training data from the origin of the coordinate system with the largest possible margin, and not two distinct classes from each other. The intent of the SVDD however is different: Instead of trying to separate the data from an origin linearly via a hyperplane, the SVDD learns a sphere encapsulating the training data. Both learning methods are shortly presented in the following sections.

\subsubsection{One-Class Support Vector Machine}

%

The objective of the One-Class Support Vector Machine is to learn a hyperplane which separates the training data from the origin of the coordinate system. Such a hyperplane can uniquely be defined by a vector $w$ orthogonal to the hyperplane, which is to be minimized in order to maximize the margin in between the training data $\lbrace x_i \rbrace_{i=1\dots N}$ and the coordinate origin. 
The primal optimization target is defined as
\begin{align}
\underset{w,b,\xi}{\min} &~\frac{1}{2}\|w\|^2 + C \sum\limits_{i}\xi_i - b \label{eq:ocsvm_prim}\\
\text{s.t.} & ~\forall i ~ w^T\Phi(x_i) \geq b - \xi_i \nonumber\\
			& ~\forall i ~ \xi_i \geq 0 \nonumber
\end{align}
with $C$ serving as a regularization parameter, trading off the maximization of the margin width $b$ and training errors $\xi_i$ caused by data points not being projected beyond the hyperplane. One usually relies on the kernelized formulation for training and evaluation, from which the prediction function Equation \ref{eq:predfunc} follows. A data point $x$ is then classified as anomalous if $f(x)<0$.
\begin{align}
f(x) = \sum\limits_i \alpha_i k(x_i,x) - b
\label{eq:predfunc}
\end{align}

\subsubsection{Support Vector Data Description}

%

The SVDD learns, given a training set $\lbrace x_i \rbrace_{i=1\dots N}$,  the smallest possible sphere enclosing that data, defined by a learned centroid $\hat\mu$ and a radius $\bar R$. The primal objective function is
\begin{align}
\underset{\hat{\mu},\bar{R},\xi}{\min} & ~\bar{R} + C \sum\limits_{i}\xi_i \label{eq:svdd_prim}\\
\text{s.t.} & ~\forall i~\|\Phi(x_i) - \hat{\mu}\|^2 \leq \bar{R} + \xi_i,\nonumber\\
			& ~\forall i~\xi_i \geq 0, ~\bar{R} \geq 0 \nonumber 
\end{align}
Similar to other support vector methods, training and evaluation is usually done in the method's dual form, with the following dual prediction function
\begin{align}
f(x) & = ~\bar{R} - a(x) \label{eq:svddpred} \\
a(x) & = ~k(x,x) - 2\sum\limits_i \alpha_i k(x_i,x) + \sum\limits_{i,j} \alpha_i\alpha_j k(x_i,x_j) \label{eq:svddpred_a}
\end{align}
Here, $a(x)$ describes the squared euclidean distance between $x$ and $\hat\mu$ in the feature space defined by the kernel function $k(\cdot, \cdot)$. A prediction point $x$ is considered anomalous if $f(x)<0$.
%

%% file: sections/behavioral_anomaly_detection.tex
\section{Behavioral Anomaly Detection}
\label{sec:behavioralanomalydetection}

The motivation for the proposed behavioral anomaly detection method is to detect anomalous behavior in the monitored components of Media Cloud applications, corresponding to the vulnerabilities described in Section \ref{sec:mediacloudvulnerabilities}. It is designed to work complimentary to the detection method for structural anomalies, described in Section \ref{sec:structuralanomalydetection}. For this purpose specific parameters of a monitored Media Cloud application need to be logged. This data can then be used to train an individual anomaly detector for each monitored component. This detector is then able to classify new communication data of this component, to detect anomalies within. The following section describes the required measurements, the feature representation and the actual application of the one class learning method for training the detector.

\subsection{Measured Values}
\label{measuredvalues}

Applications running in cloud infrastructures need to be highly optimized for the cloud principles of flexibility and scalability. The applications and components developed for the the Media Cloud conform to those principles. In terms of monitoring for security purposes, such applications are harder to track, because the components are dynamically distributed between different computational resources. For this reason, the proposed method does not monitor the behavior of the computational resources or of the applications, but instead it focuses on the behavior of the measured values at individual components. As elaborated above, all applications are built using a set of software components. When used in another instance of the same application type, a component is expected to behave similar for each instance of this application type. The same component may however also be used in a different application type, which results in a different behavior in instances of this different application type.

\begin{table}[ht]
\vspace{-4mm}
\centering
\begin{tabular}{ c | l}
\textbf{$V_t$} & Description \\
\hline
\hline
$v_1$ & Avg. CPU time (processes per min)\\
\hline
$v_2$ & Avg. Num. of Received Msgs. per sec\\
\hline
$v_3$ & Avg. Num. of Sent Msgs. per sec \\
\hline
$v_4$ & Avg. Size of Received Msgs.\\
\hline
$v_5$ & Avg. Size of Sent Msgs.\\
\hline
\end{tabular}
\vspace{4mm}
\caption{The values $V_t$ measured on each monitored component.}
\vspace{-4mm}
\label{tab:componentmeasurements}
\end{table}

Thus the behavior of all components of each application instance need to be monitored. A dictionary stores the type of application per application instance. This allows a consolidation of the data of all instances of the same application type during run-time.
For each component a tupel of measured values $V_t = \{v_1, v_2, \dots , v_5\}$ is logged in a selected time interval (e.g. one second), which allows a fine grained monitoring of the behavior of the individual components. The values where selected to be able to represent maliciously behaving communication, e.g. in terms of anomalous message rates or message sizes. Table \ref{tab:componentmeasurements} describes those measured values. An increasing number of received messages per second ($v_2$) might for example point to a DoS-attack.

\subsection{Feature Representation}
\label{featurerepresentation}

The selected learning method requires a vector representation of the data for training and prediction. Although sparse feature representations have been evaluated as well, and were promising for their special properties (e.g. encoding the temporal development into a spatial representation), a dense feature representation has shown the overall best performance. The behavioral anomaly detection method is designed to effectively make use of the temporal development of the measured values. Therefore the feature mapping is based on a projection function $m(t)$ over a sliding window of size $s$. For each time $t$ there are measured values $V_t = \{v_1, v_2, v_3, v_4, v_5\}_t$ available. The projection function utilizes $V_t$ and the $s-1$ previous datapoints $\{V_{t-1}, \dots, V_{t-(s-1)}\}$ to map the measured values onto a new $5s$-dimensional data vector, containing information about the temporal development within this sliding window. This projection defines a new data point $x_t$ at time $t$ as 
\begin{align}
& x_t = m(t) = \{v_{i,t}\} \\
&\forall i \in \{1, \dots, 5\}, t \in \{1, \dots, s-1\} \notag
\end{align}
i.e.
$$
m(t) =\{\underbrace{v_1, \dots , v_5}_{V_t}, \;
\underbrace{\dots}_{V_{t-1}}, \; \dots, \underbrace{\dots}_{V_{t-(s-2)}}, \;
\underbrace{v_1, \dots , v_5}_{V_{t-(s-1)}}\}
$$

The motivation behind these features is to model the general dynamic behavior of the different values $V$, because attacks to the Media Cloud system, as those described in Section \ref{sec:mediacloudvulnerabilities}, would cause the values to behave differently than usual, for a longer period of time.

\subsection{Prediction of Anomaly Scores}
\label{sec:behav_model_and_prediction}
Although Section \ref{sec:learningmethods} presented two different methods for using support vector machines for anomaly detection, in case of the behavioral anomaly detection method it is sufficient to use the One Class SVM method. We are using gaussian radial basis function kernels, defined as follows

\begin{align}
k(x,x') &= \exp \left( \gamma \| x-x' \|^2_2 \right)\\
\notag \gamma &= -\frac{1}{2\sigma^2}
\label{eq:rbfkerndef}
\end{align}

because the resulting models are proven to be equivalent to the smallest enclosing high-dimensional hypersphere around the provided data points, as defined in the SVDD-approach by Tax \cite{tax2004support}, which means evaluating the One Class SVM method is sufficient in this case. As described in Section \ref{sec:oneclasslearningmethods}, the One-Class SVM learns a model for a hyperplane separating the projected data from the point of origin. 
As discussed in Section \ref{measuredvalues}, the behavioral anomaly detection
focuses on learning a model based on specific extracted features of the data for each combination of components and application class. The final one-class model describes those sets of normal data. For the classification of a new data point $x$ at a time point $t$, denoted as $x_t$,  the function $f(x_t)$ calculates an anomaly score, which is based on the distance of $x_t$ from the hyperplane, i.e. 
\begin{align}
f(x_t) = \sum_i \alpha_i k(x_i, x_t)
\end{align}
where $k(x_i, x_t)$ is the RBF kernel function and $\alpha_i$ are the learned support vector weights.  
To predict, whether $x_t$ is normal or anomalous, the threshold $b$ (introduced in equation \eqref{eq:predfunc}) on the anomaly score is defined for each model. The threshold is calibrated on a validation data set, with the objective to achieve the lowest possible false positive ratio. The actual prediction $p(x_t)$ is then conducted with the function 
\begin{align}
p(x_t) = sign(f(x_t)-b)
\end{align}
where $x_t$ is anomalous if $p(x_t)<0$ and otherwise normal. In practice, each component of each Media Cloud Application has to provide such training and validation data. Once the anomaly detector of each component is trained, the detector is coupled with this component, monitoring its behavior and raising an alarm, if an anomaly is detected.

%% file: sections/structural_anomaly_detection.tex
\section{Structural Anomaly Detection}
\label{sec:structuralanomalydetection}
The structural anomaly detection method is designed to work complimentary to the detection method for behavioral anomalies, described in Section \ref{sec:behavioralanomalydetection}. Instead of using timelines of specific measured values, it learns a model for each application, based on logs of the dynamically changing application graph, i.e. the interconnected components within each application, to detect anomalies visible in this graph structure, as described in Section \ref{sec:mediacloudvulnerabilities}. Once the model is learned, anomalous components and anomalous connections between components can be detected. Furthermore the anomaly can be located, due to the use of additive kernels and a sum-pooled explicit feature representation for retracing the origins of the classification decision.
The following sections elaborate the feature presentation, the actual prediction of anomalies and the methods used to locate the predicted anomalies within the graph.

\input{sections/graph_features}

\input{sections/local_prediction}

%% file: sections/graph_features.tex
\subsection{Feature Representation}

Our data consists of structural snapshots of the application at times of measurement in form of highly-interconnected graph structures. We use standard graph notation, describing our graph structures $G=(V,E)$ as a pair of labeled nodes $V$ and directed edges $E \subseteq V\times V$.
The concept of Bag of Substructures is closely related to the Bag of Words representation of documents used in natural language processing: In essence, a BoS representation counts numbers of occurrences of discrete subunits present within an input graph structure. Let $G$ denote a graph structure to be projected into the BoS feature space and $s$ denote a substructure contained within $G$, consisting of a set of nodes $V_s$ and edges $E_s$. The calculation of a BoS feature is then defined as
\begin{align}
x & = \sum\limits_{s \in G} \mathds{1}_{m(s)} \label{eq:bosmap}
\end{align}
where $m(s)$ describes a mapping function relating the substructure $s$ to a dimension $d$ in BoS feature space. The final BoS representation of $G$ is then a sum-aggregation of all binary indicator vectors $\mathds{1}_d$ which contain an entry $1$ in dimension $d$ and $0$ otherwise. 

The class of substructures we consider are degree-$1$-neighborhoods around each node within an input graph. The neighborhood of 1st degree around a selected node $u$ contains all nodes $v$ which share a direct edge connection with $u$, regardless of the direction of those edges, as well as the node $u$ itself. Those nodes $V_s = u\ \cup \lbrace v \rbrace$, together with \emph{all} edges $E_s = \lbrace \lbrace v' , v \rbrace | \lbrace v' , v \rbrace \in E\ \land\ v' \in V_s\ \land\ v \in V_s \rbrace$ connecting the nodes from within $V_s$ to capture clique-structures as well, define our neighborhood substructures $s = (V_s, E_s)$. By ignoring edge directions while expanding the neighborhoods, and due to the extraction of one neighborhood for each node within the graph, information about missing connections can be encoded into the feature vector and is therefore locatable later on. 

Our choice of $1$st degree neighborhoods is grounded on their range of expressiveness. They fully include sequences and arbitrarily connected substructures containing up to three nodes, and they partially include larger structures, especially around highly-interconnected cliques and nodes with high edge degrees. Other than the degree of neighborhood expansion, the substructures are not restricted in size and are therefore capable of describing forks within the graph structure, as well as the cardinality thereof. Another argument is their low computational complexity: Calculating the full set of substructures for a graph $G$ can be done in $\mathcal O(n)$, with $n$ being the number of nodes in $G$. Using appropriate data structures for $G$ and hashing functions for $m(\cdot)$ (e.g. Liu et.al. \cite{LiuWanKumCha11}), mapping a graph into a vector space is near-instantaneous -- an important property for monitoring a system in real-time. We found binarized and normalized features to perform best with both considered kernel methods.



%% file: sections/local_prediction.tex
\subsection{Prediction and Localization of Anomaly Scores}
\label{sec:pred}

Both kernel methods described in Section \ref{sec:learningmethods} use a thresholded prediction function $f(x)$ to determine a prediction point $x$ as normal or an outlier.
The respective thresholds can either be learned \cite{scholkopf1999support,tax2004support} or set manually to satisfy given demands on available normal data. We follow the latter, more practical approach, which allows us to adapt the classifier to the required use-case. In day-to-day operation with numerous instances of multiple application types active in the cloud, the classification threshold is set to a low false positive rate, to raise alarms only in the most severe situations. Once an in-depth monitoring of critical applications is required, higher false positive rates are acceptable in order to increase the sensitivity of the classifier.

While the prime objective of our system is the detection of anomalous substructures in application graphs, we are also very interested in the exact location of this substructure, i.e. we wish to rate nodes and edges individually according to their contribution to the graph's prediction value.
In case a graph is completely structurally unrelated to the data previously trained on, the cause of the anomaly rating is evident. However, as in \cite{eberle2007mining}, we expect most anomalies to be local phenomena, affecting only small subsets of nodes and edges, whereas the majority of the graph structure remains true to the expectation of the learned model. Considering the potential dimensions of the non-planar MediaCloud application graphs, counting well above 150 highly interconnected components, a localized threat assessment in order to reduce the processing time for the administrators is extremely relevant.

Due to the Bag of Substructures feature extraction we know that each dimension in BoS space corresponds to a unique structural confirmation of nodes and edges. By restricting our choice of kernel functions to the family of sum-decomposable kernels for training and classification, the value produced by the prediction functions $f(x)$ can be naturally decomposed into differential contributions for each dimension $d$ of the prediction point $x$.
We regard a kernel function as sum-decomposable, if an equivalent formulation based on kernel functions acting on single input dimensions exists, i.e. such that the kernel function has the property
\begin{align}
k(x,x') = \sum\limits_d k(x^{(d)},x'^{(d)})
\label{eq:addkerndef}
\end{align}
with $x^{(d)}$ describing the value of dimension $d$ in data point $x$.
Two kernel functions meeting the above criterion have been used for our method, namely the histogram intersection kernel and the linear kernel. A decomposition of the linear kernel in accordance to equation \eqref{eq:addkerndef} is
\begin{align}
k(x, x') = \langle x , x' \rangle = \sum\limits_d x^{(d)}x'^{(d)}
\end{align}

For the histogram intersection kernel the required decomposition is already provided via the kernel function's definition:
\begin{align}
k(x, x') = \sum\limits_d \min(x^{(d)},x'^{(d)})
\end{align}

In the remainder of this section we will discuss a decomposition of the prediction functions $f(x)$ into scores for each substructure extracted from a graph and finally anomaly ratings for individual nodes and edges. The proposed decomposition is calculated as a series of consecutive steps and can be understood as an inversion of the classification pipeline from the data point as a graph structure towards the prediction of the classifier. The resulting localized threat indication for graph substructures can then be visualized in order to allow a quick identification of the sources of structural error.

\subsubsection{Dimensional Prediction: SVDD}

The prediction function \eqref{eq:svddpred} of the SVDD consists of the learned radius $\bar{R}$ minus a function $a(x)$ describing the squared euclidean distance between the prediction point $x$ and the learned centroid $\hat{\mu}$. In order to identify the local sources of the graph's anomaly score we ignore the thresholding in $f(x)$ and concentrate on the decomposition of function $a(x)$, describing the distance between the learned centroid $\hat\mu$ and the point $x$, into dimensional contributions $g(x)^{(d)}$ of single input dimensions. 
Using sum-decomposable kernels as in equation \eqref{eq:addkerndef} yields for each dimension $d$ the predicted value
\begin{align}
g(x)^{(d)} &= k(x^{(d)},x^{(d)}) - \sum\limits_i \alpha_i k(x_i^{(d)}, x^{(d)}) + \sum\limits_{i,j} \alpha_i\alpha_j k(x_i^{(d)},x_j^{(d)})
\end{align}

\subsubsection{Dimensional Prediction: One-Class Support Vector Machine}

Similar to the case of the SVDD, we are interested in the origins of the prediction and do again ignore the additive bias term.
In contrast to the prediction of the SVDD, the OCSVM, without the thresholding,  describes the benignity of a prediction point $x$: The better a vector $x$ describes the direction $w$ orthogonal to the hyperplane, the higher the predicted value.
Dimensional predictions $g(x)^{(d)}$ yielding low values therefore indicate a lack of benignity whereas higher values represent expected graph structures. We use the definition in equation \eqref{eq:addkerndef} in conjunction with an arbitrary sum-decomposable kernel function to obtain dimensional scores
\begin{align}
g(x)^{(d)} = \sum\limits_i \alpha_i k(x_i^{(d)}, x^{(d)})
\end{align}

\subsubsection{Local Score Projection}
\label{sec:localscore}

We have at this point calculated predictions for each dimension of $x$. Our next objective is the assignment of those scores to groups of nodes and edges matching the substructures representing the predicted dimensions in BoS space.
To each substructure $s$ contained within $G$ and considered by the feature extraction scheme, we assign a fraction $q(s)$ of the prediction $g(x)^{(m(s))}$. Here $m(s)$ is the same mapping function as used in equation  \eqref{eq:bosmap}, relating a structure $s$ to the index of a dimension in BoS space.
To take the sum-aggregation used in equation \eqref{eq:bosmap} into account we define that fraction to correspond to the weighted contribution of $s$ to $x^{(m(s))}$ during the BoS feature calculation:
\begin{align}
q(s) = \frac{g(x)^{(m(s))}}{x^{(m(s))}}
\end{align}

The substructure scores $q(s)$ are then mapped onto compatible node- and edge-configurations by averaging, to consider varying coverage of different parts of the graph by the scope of the extracted substructures. This results in predictions $r(o)$ for all node and edge objects $o \in G$:
\begin{align}
r(o) = \frac{1}{|S|}\sum\limits_{s\in S} q(s) ~;~ S = \lbrace s | o \in s\rbrace
\end{align}
The set $S$ describes all extracted substructures covering the current node- or edge-object $o$. For models obtained using an SVDD the calculations of local predictions are finalized at this point. For the sake of consistency we convert the local benignity ratings obtained in case of the OCSVM into anomaly ratings, or rather a \emph{lack of benignity}, by assigning to each $o$ the difference between the score of the most benign node or edge and $r(o)$.
\begin{align}
r(o)_{mal} =  - r(o) + \max \lbrace r(o) \rbrace_{\forall o \in G}
\end{align}

Prior to visualization local prediction scores $r(o)$ are scaled such that $r(o) \in [0,1]~\forall o$. Depending on the mode of operation one can consider either all $o$ within the scope of a single graph, or alternatively a set of graphs. We normalize as
\begin{align}
r'(o) = \frac{r(o)}{\max \lbrace r(o) \rbrace_{\forall o \in G}}
\end{align}
using $G$ here either as a single graph or a combination of  multiple graphs. Note that due to exclusion of the thresholding from both classifiers and the conversion to anomaly ratings in case of the OCSVM all prediction scores $r(o)$ and $r'(o)$ are always larger than or equal to zero. We visualize the predictions by linearly mapping the (normalized) prediction scores to some color map of choice.

%% file: sections/datasets.tex
\section{Data sets}
\label{sec:datasets}
For testing both proposed methods, a realistic Media Cloud video conference application has been executed multiple times, each run having a length of 10 minutes. In each run the values $v_t$ have been logged with a sampling rate of 1 sec. Two sets of data have been created. The first set of runs, aimed at testing the behavioral anomaly detection, focuses on a higher variance in the amount of background processes, resulting in multiple runs consisting of up to 57 components and 18000 events on average. The second data set, aimed at the structural anomaly detection, focuses on the number of connected devices, and does therefore incorporate up to 160 components, while the number of tracked events is only 6112. This specific data set also incorporates logs of the dynamically changing graph structure.


\input{sections/datasets_behavioral_data}

\input{sections/datasets_structural_data}


%% file: sections/datasets_behavioral_data.tex
\subsection{Data sets for behavioral anomaly detection}
\label{subsec:datasets_behavioral}
\begin{figure}
\vspace{-4mm}
\subfigure[$V$ of a single component. ($v_5=1$).]{
\includegraphics[width=0.50\textwidth]{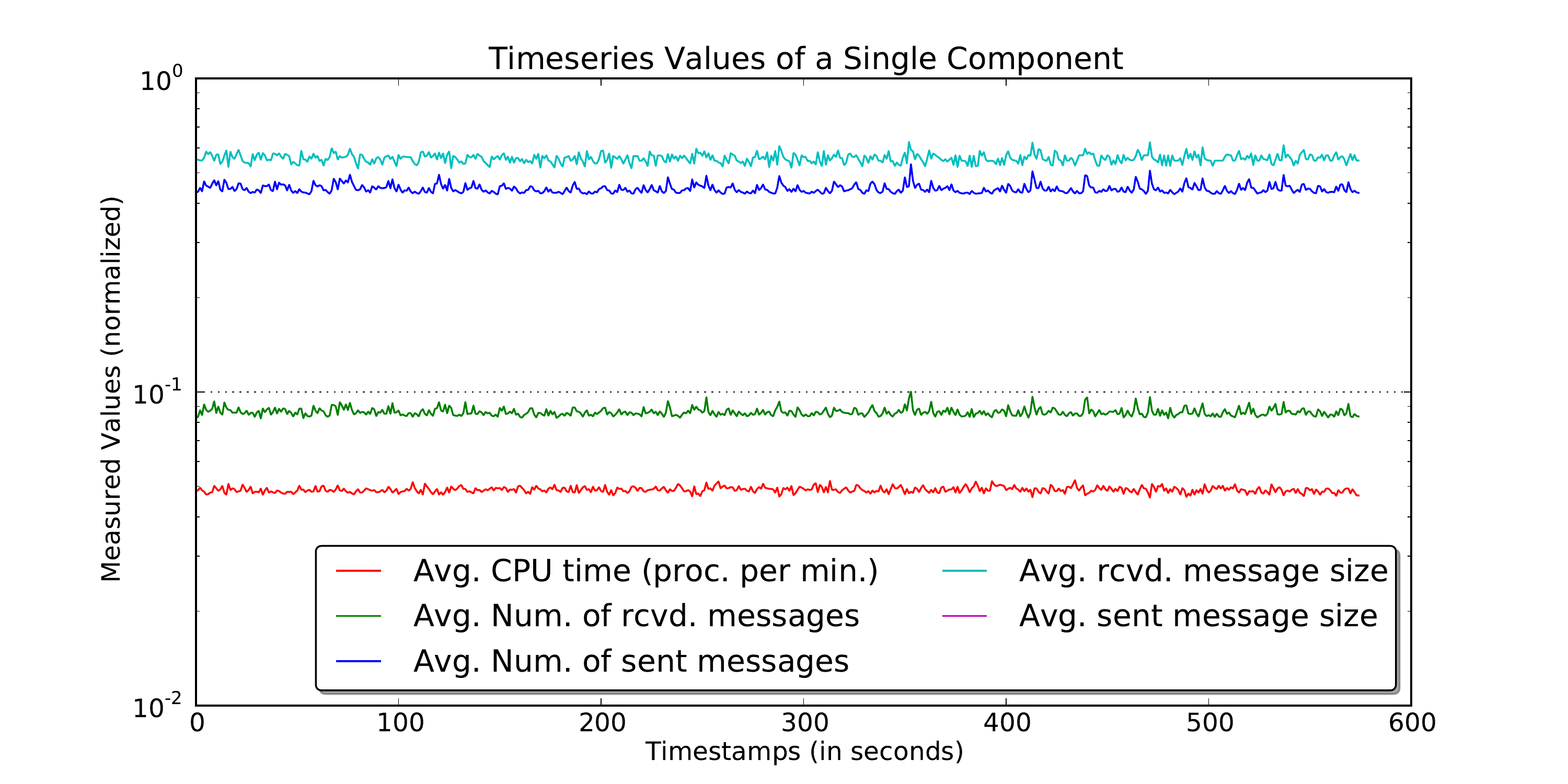}
\label{fig:alumeasuredvalues}
}
\hfill
\subfigure[Examples of dynamic behavior.]{
\includegraphics[width=0.50\textwidth]{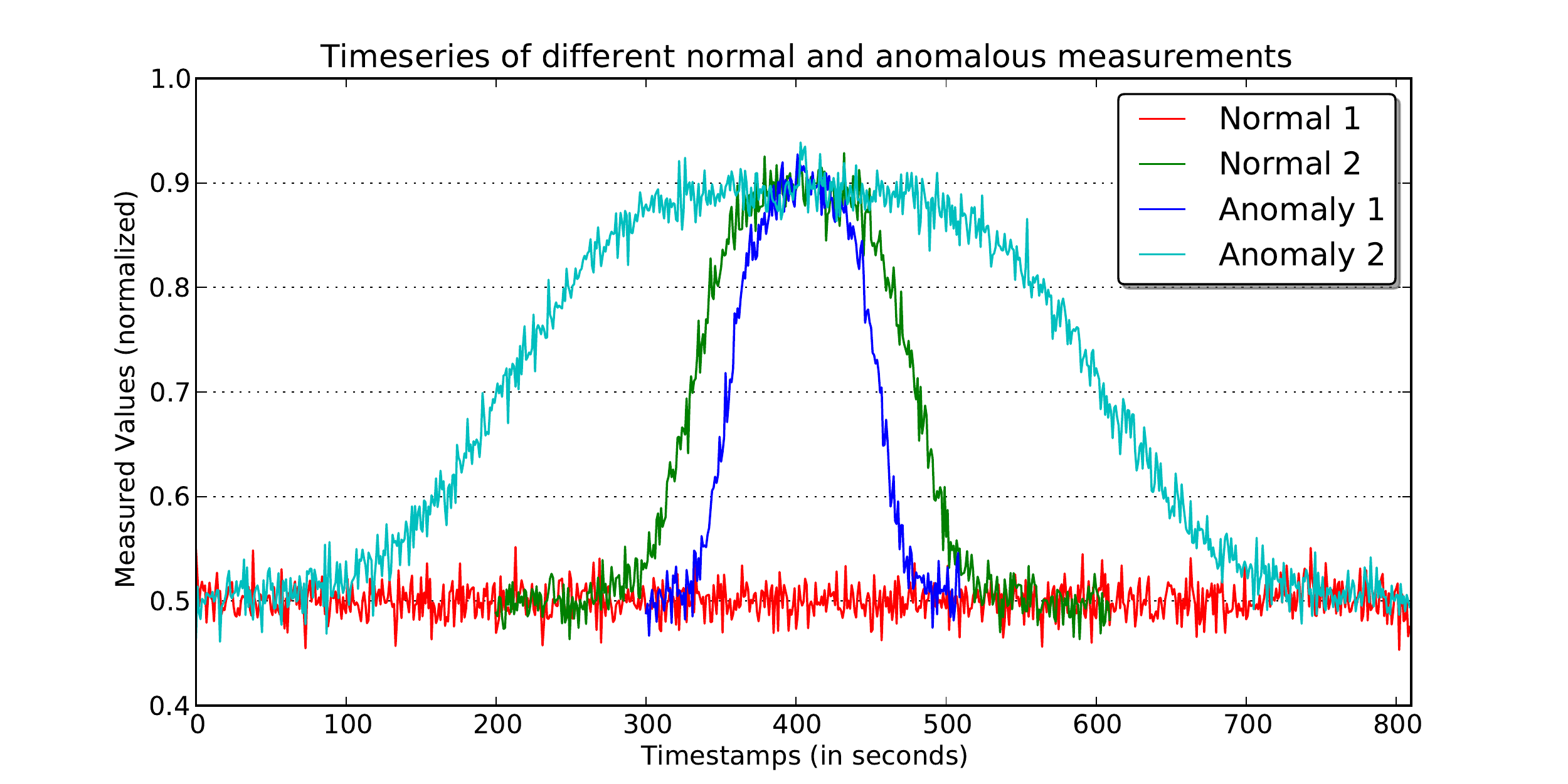}
\label{fig:datasetsAndAnomalies}
}
\caption{Different examples of measured timeseries.}
\vspace{-2mm}
\end{figure}

The measured values of the components show only a very small variance. Figure \ref{fig:alumeasuredvalues} shows the normalized values for the single most active component, a central video-encoding component. The values show an overall small variance and no specifically dynamic behavior like ascending or descending sections within. There is also no anomalous data available in those logs, to empirically evaluate the detection performance of the behavioral detection method. To address this problem, the real data has been used to craft a more dynamic version, as well as an anomalous data set.

To achieve a more dynamically behaving time series, the timelines of the real data are combined with parametric sigmoid functions, allowing for smooth dynamic gradients. To achieve more variance in the anomalous data, random gaussian noise has been added there. The required variance is derived from the statistics of the video conference data. The dynamic behavior realized by the sigmoid functions follows a Media Cloud paradigm, which allows the activity of application components to be dynamically decreasing or increasing during run-time - a behavior unfortunately not occurring in our data set. Figure \ref{fig:datasetsAndAnomalies} depicts examples of different simulated time series measurements of a single $V_t$. The values are normalized and show a non-dynamic (\textit{Normal 1}) timeline with constant behavior, as found in the video conference data, and a dynamic version (\textit{Normal 2}) of normal data simulating a dynamic increase and decrease of the measured value, as it may occur when a specific activity of this component is triggered. The plot \textit{Anomaly 1} represents a relatively short burst of anomalous activity, as it may be perceived due to data breaches or malfunctioning components, while \textit{Anomaly 2} represents a much wider temporal range of activity, perceivable in the case of traffic hijacking or DoS-attacks.

%% file: sections/datasets_structural_data.tex
\subsection{Data sets for structural anomaly detection}
\label{sec:datasets_structural}

\paragraph{Media Cloud Video Conference Data Set}
results from recording a real video conference session with up to 6 concurrent users. It consists of 14 unique graph structures captured over the time, representing structural changes within the application. In this scenario a change in the application structure occurs whenever a participant joins or leaves the session. Each user is featured within the application graph by a core structure of connected components, as well as a dynamically changing chain of components reserved to connect to and communicate with the respective counter parts of other present users. In order to build a realistic yet larger training set we have drawn random graphs from the recording 400 times. 
This data set features average node counts of 80.1 and average edge counts of 119.2 per graph, and an average node degree of 2.9 per node and up to 160 components connected with 256 edges for the largest graph structure.

\paragraph{Synthetic Data Set}
has been created in order to create a richer data set in the sense of size, complexity and number of unique graphs. This has been done by using the recorded graph instances, manually analyzing their structural patterns and using a generator to construct graph structures adhering to those rules. This data set features average node counts of 119.2 and average edge counts of 194.1 per graph, as well as an average node degree of 3.3 per node, with the largest graph consisting of 516 components connected with 924 edges.

%% file: sections/evaluation.tex
\section{Evaluation}
\label{sec:evaluation}
The evaluation of the methods for behavioral and structural anomaly detection is conducted on the data sets described in Section \ref{sec:datasets}. Since each method focuses on different aspects of the data set, each method is evaluated individually.

\input{sections/eval_behavioral}

\input{sections/eval_structural}

\input{sections/evasion}

%% file: sections/eval_behavioral.tex
\subsection{Behavioral Anomaly Detection}
\label{evaluationbehavioralanomalydetection}

For all experiments the behavior of the ten most active components has been used. The average number of samples was 570. Since each application run has been executed multiple times, the data has been separated to obtain individual data sets for training, validation and testing. Finally the results presented in the following section are averaged over the ten different components. To achieve an optimal representation of the behavioral characteristics over time, the parameter $s$, responsible for the size of the sliding window, is of crucial relevance. The experiments described below are conducted using a range of values for $s \in [20,160]$. Shorter values do not represent the temporal development adequately. Larger values, on the other hand, are not required, as they do not further improve the results. Analyzing the range of $s$ allows deeper insights into the temporal requirements and characteristics of the simulated Media Cloud data, and allows conclusions about ways to handle them in practice.

\begin{figure}[htb]
\vspace{-4mm}
 \centering
 \includegraphics[width=0.51\textwidth]{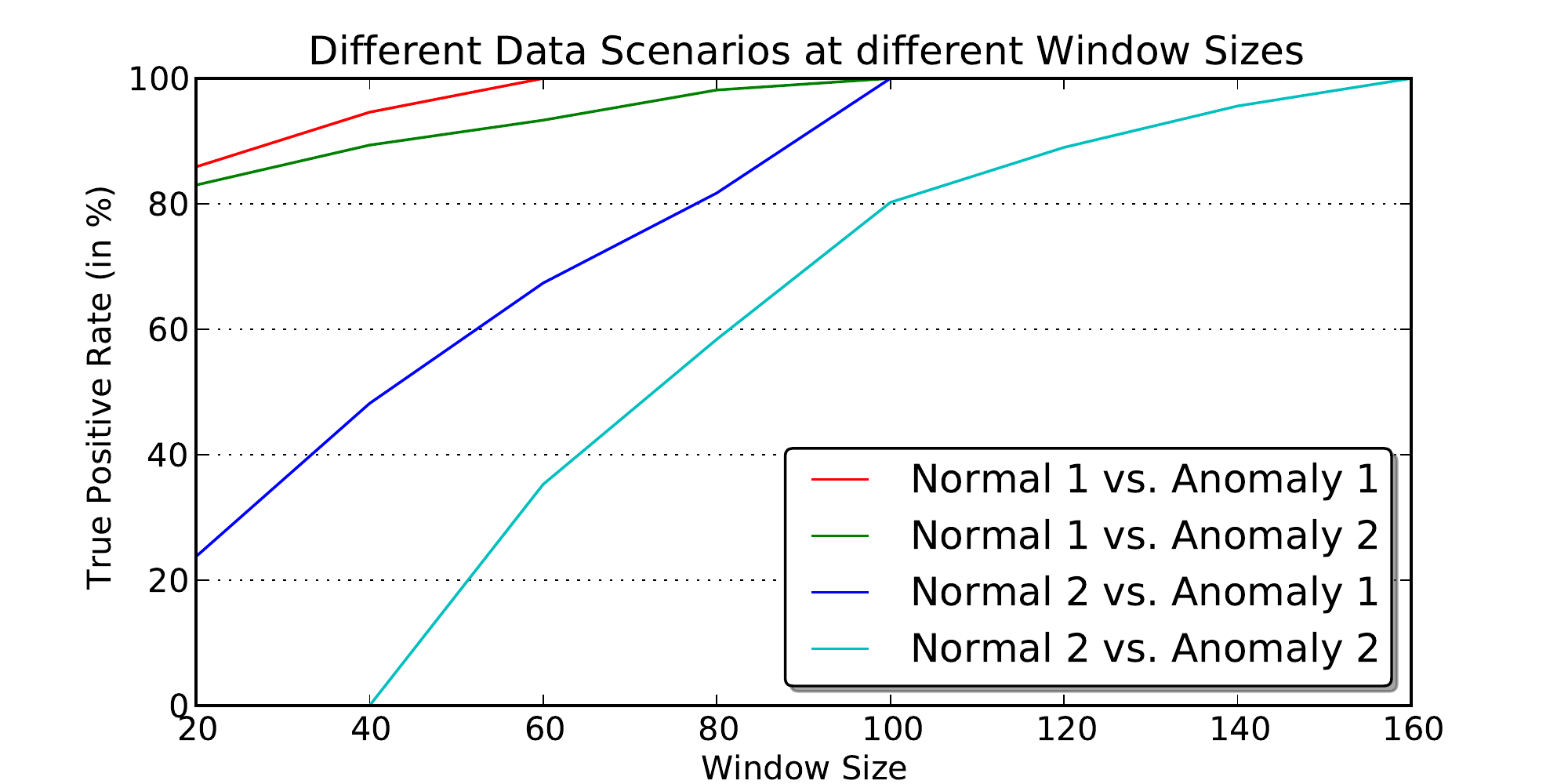}
 \caption{Average True Positive rates at different window for the four experiments.}
 \label{fig:windowsizetprates}
\end{figure}

In all of the following experiments, depicted in Figure \ref{fig:windowsizetprates}, the false positive rate is well below $1\%$, and for the optimal models even at $0\%$. The achieved true positive rates are much more interesting. The first experiment, \textit{Normal 1 vs. Anomaly 1} utilizes the correspondingly modified distributions, depicted in Figure \ref{fig:datasetsAndAnomalies}, i.e. no dynamical behavior is added to the video conference data. The resulting model is then tested against other runs of \textit{Normal 1} as well as data modified according to \textit{Anomaly 1}., The true positive rate starts at a value of $86\%$ at $s=20$ and approaches its optimum of $100\%$ already at $s=60$. In comparison with the other experiments, this is the smallest optimal value for $s$. The second experiment based on the non-dynamic normal data is the one of \textit{Normal 1 vs. Anomaly 2}. This time the anomalous behavior covers a wider duration, and requires also a larger window size of $s=100$ to be detectable at $100\%$. Both of those results are rather intuitive, because they represent the simplest experiments possible, based on a model of nearly constant normal data which deviates strongly from the applied anomalies. In fact it is so simple, that a threshold-based baseline detector would have been capable of an identical, if not better performance.
Before discussing the experiments which are based on the dynamic time series \textit{Normal 2}, it is important to note that none of the different applied window sizes is large enough to represent any of the dynamical curves as a whole in a single feature. Instead, only subsections of the curve are represented as individual features.
In the experiments based on the data modified according to \textit{Normal 2}, which shows a specific dynamic behavior, strongly deviant from the one of \textit{Normal 1}, the results are much more interesting. Because the normal data already contains decreasing and increasing gradients, a simple threshold-based baseline detector is not able to differ the normal data \textit{Normal 2} from the anomalous time series \textit{Anomaly 1} and \textit{Anomaly 2}. It could be applied if the peak of the anomalous data would deviate from the peak of the normal data, enabling at least the detection of this deviation. But this is intentionally not the case here. The proposed method, however, is still capable of a detection rate of $100\%$. And while \textit{Anomaly 2} requires a value of $s=160$ to reach its optimum, the detection of the narrower \textit{Anomaly 1} works at the previously favored value of $s=100$. Those results show that the proposed method allows the detection of anomalies with varying and dynamical behavior, even if the underlying normal data is much more dynamic than the current data logs suggest.

On both sets the detection performance achieved on \textit{Anomaly 2} was generally better with a larger window. To enable the detection of anomalies which require different window sizes, it is therefore recommended to take the largest value that is still practically feasible. The practical feasibility strongly depends on the scenario the proposed detection method is used in. If an administrator is actively analyzing a running Media Cloud system, larger window sizes help to find all kinds of anomalies, at the cost of the temporal delay of such a large window. If on the other hand immediate alarms are required during run-time, a shorter window size is advised, at the cost of a sub-optimal detection of anomalies with a longer duration. It is also useful to analyze the normal data, because in the case of a low degree of dynamic behavior, a smaller window size is sufficient.

%% file: sections/eval_structural.tex
\subsection{Graph Anomaly Detection}
\label{graphanomalydetection}

For both data sets, four experiments have been conducted, targeting one kind of anomaly similar to those considered in \cite{eberle2007mining} -- \emph{removed} nodes or edges , \emph{added} nodes or edges at unexpected locations, \emph{changed} structures as redirected edges or an inverted sequence of nodes which is a combination of the former two anomalies, as well as a combination of all mentioned kinds (\emph{mixed}). For each data set and anomaly category one training set as in Section \ref{sec:datasets_structural} and one evaluation set has been generated. The same procedure has been used for the evaluation and for the test set, followed by a modification of around 10\% of the benign graphs according to the anomaly category, such that a given edit distance (5 for the recorded data, 8 for the synthetic data) is not exceeded.
\begin{figure}[ht]
\vspace{-5mm}
 \subfigure[Recorded Data: graph-wise prediction]{
 \includegraphics[width=0.5\textwidth]{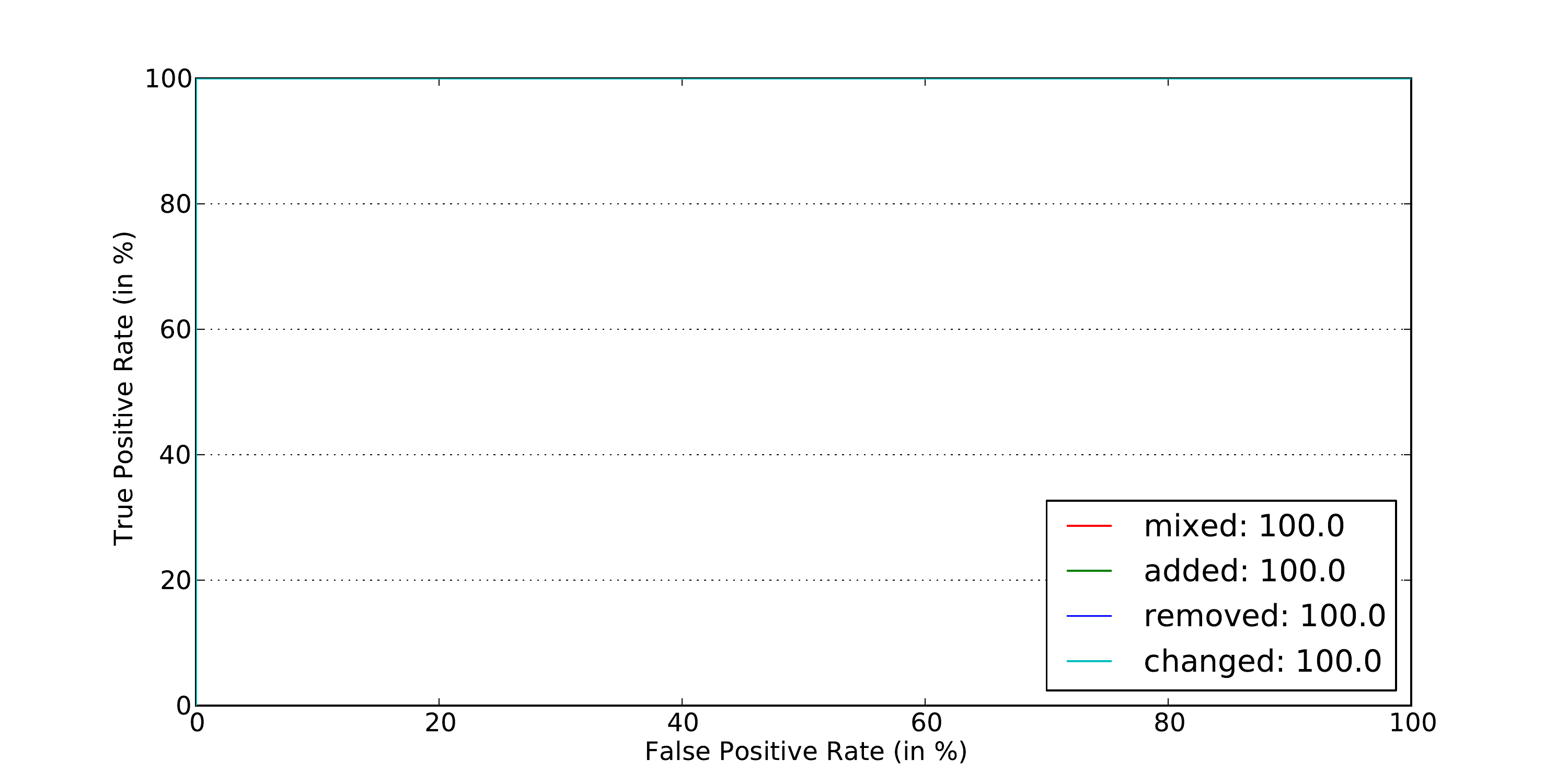}
 \label{fig:real_graphwise}
 }
 \hfill
 \subfigure[Recorded Data: local prediction]{
 \includegraphics[width=0.5\textwidth]{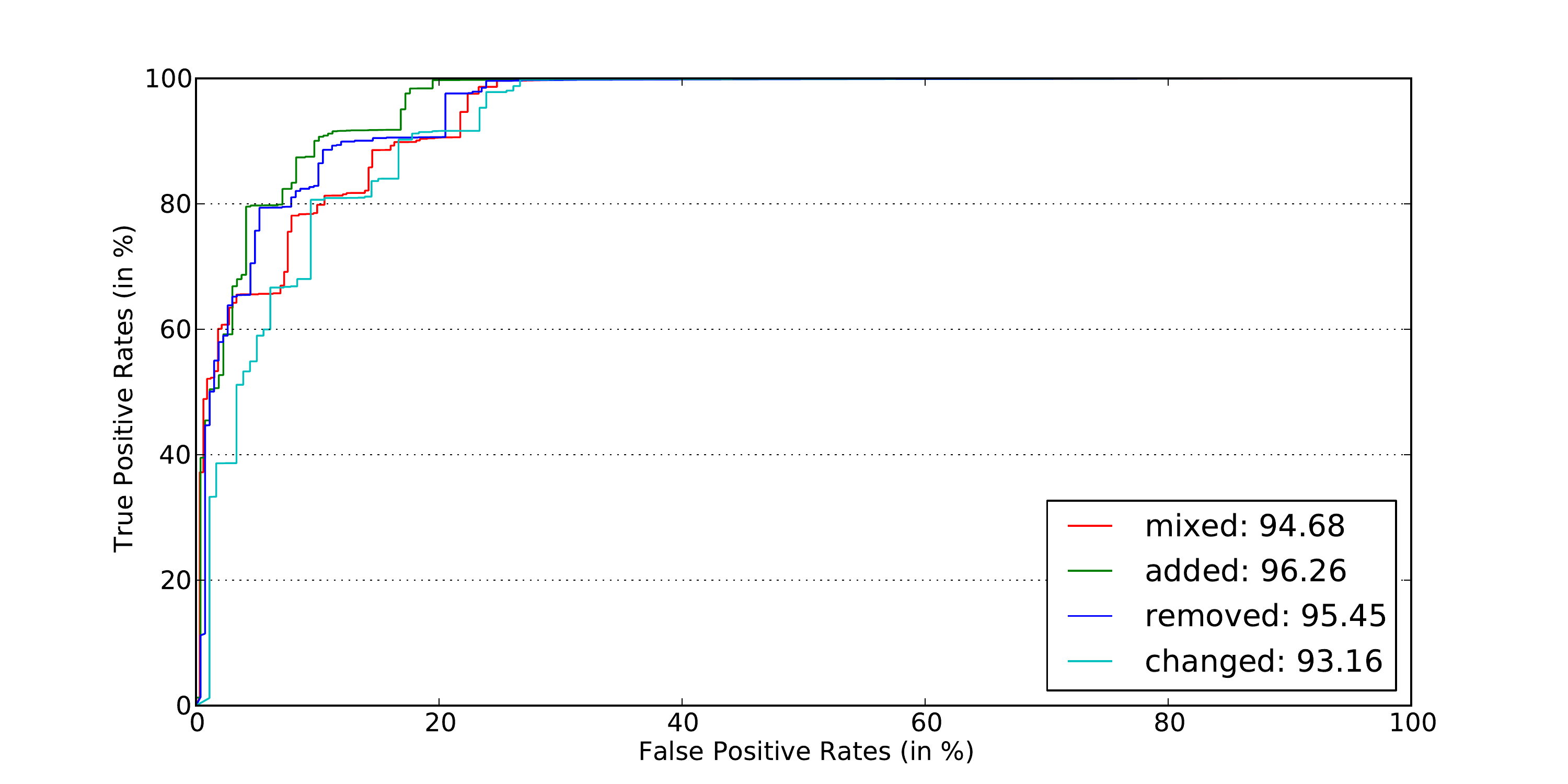}
 \label{fig:real_local}
 }
 \vspace{-3mm}
  \subfigure[Synthetic Data: graph-wise prediction]{
 \includegraphics[width=0.5\textwidth]{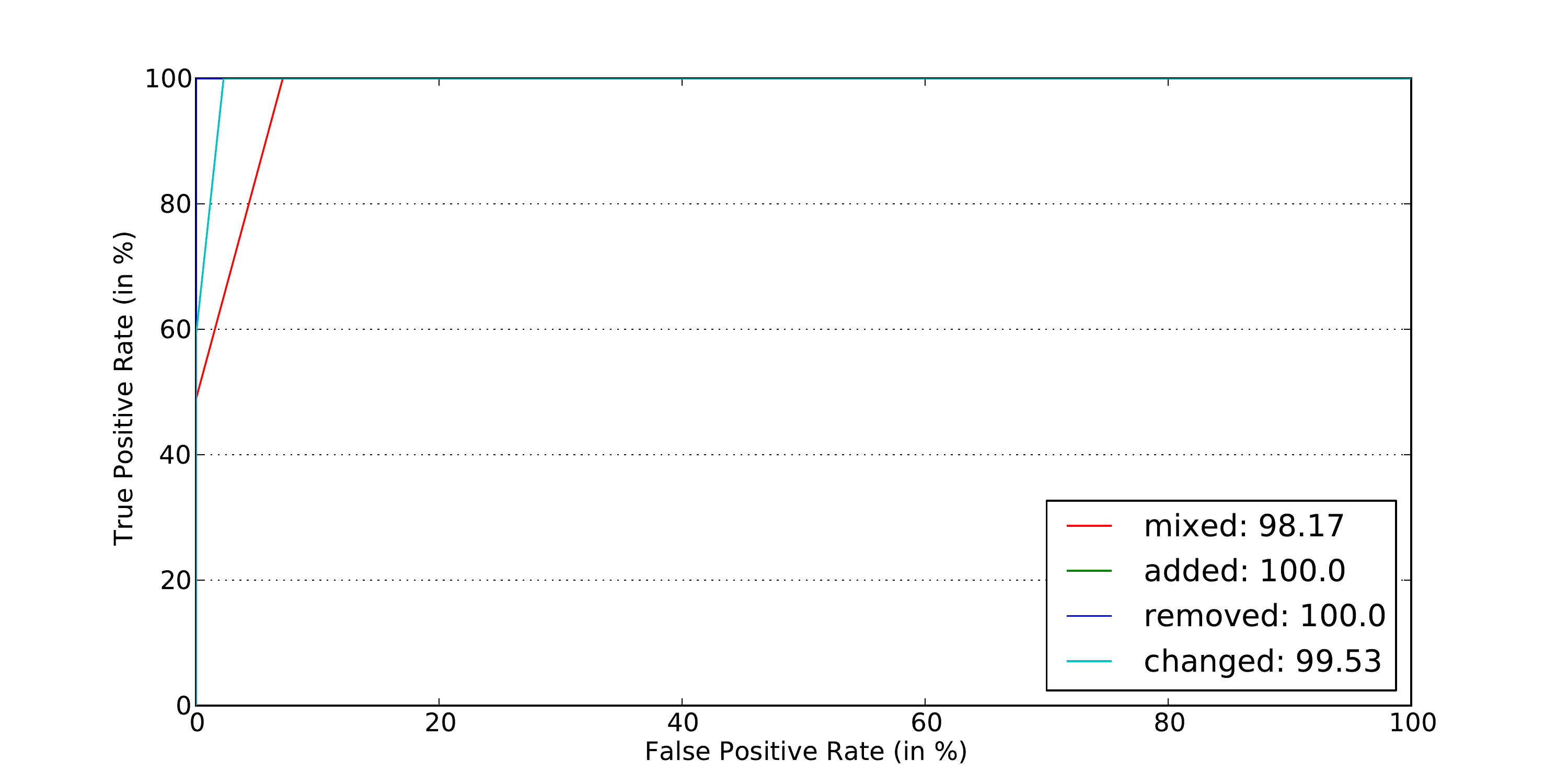}
 \label{fig:synth_graphwise}
 }
 \hfill
  \subfigure[Synthetic Data: local prediction]{
 \includegraphics[width=0.5\textwidth]{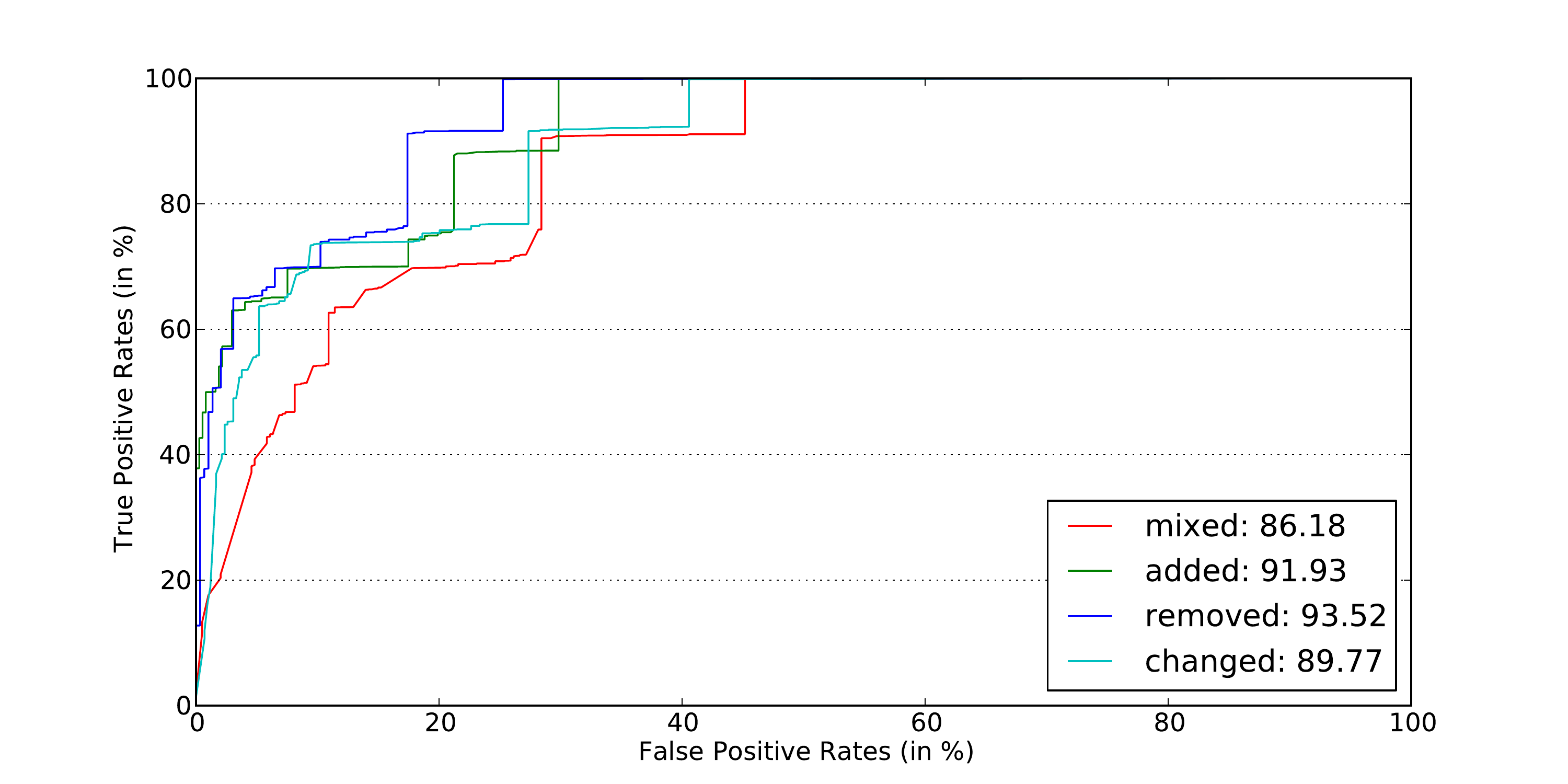}
 \label{fig:synth_local}
 }
 \caption{ROC-curves and AUC-values for graph-wise and local predictions for the real recordings and synthetic data.}
  \label{fig:struct_rocs}
  \vspace{-3mm}
\end{figure}

\begin{figure}[h]
\vspace{-2mm}
  \subfigure{
 \includegraphics[width=0.4\textwidth]{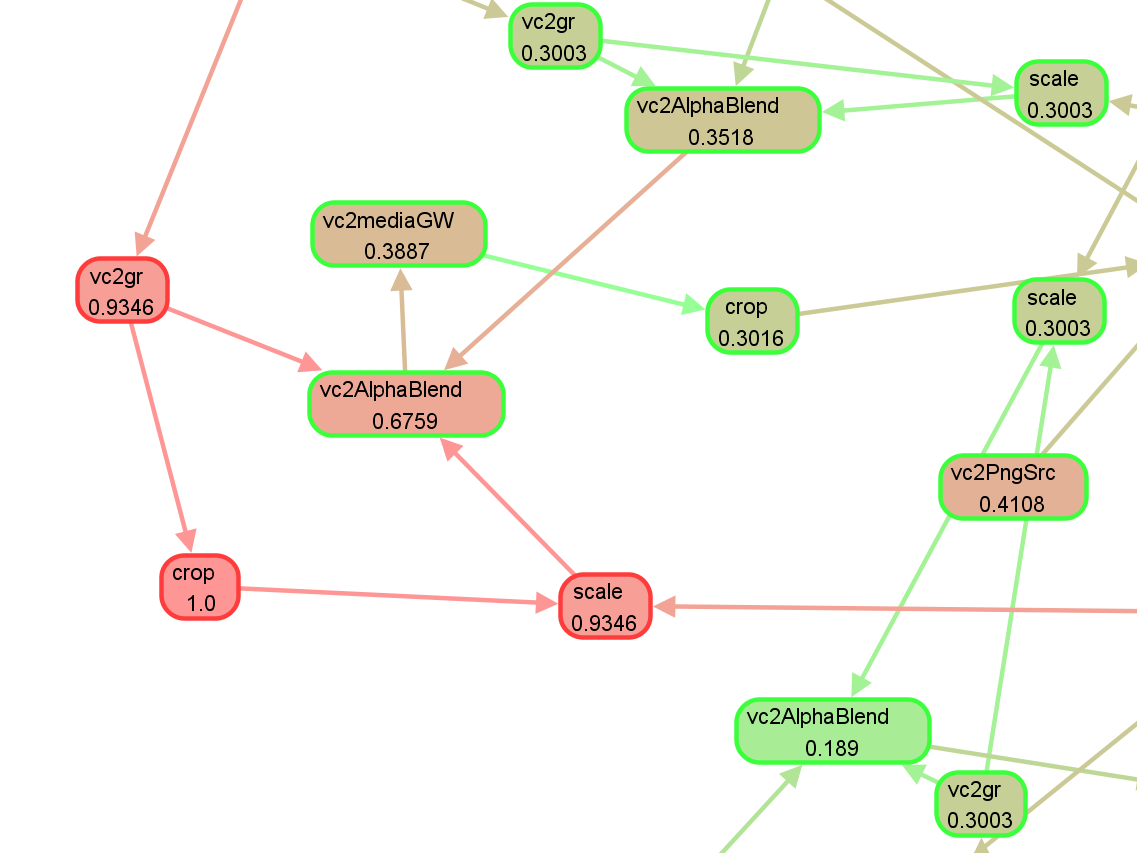}
 }
 \hfill
  \subfigure{
 \includegraphics[width=0.4\textwidth]{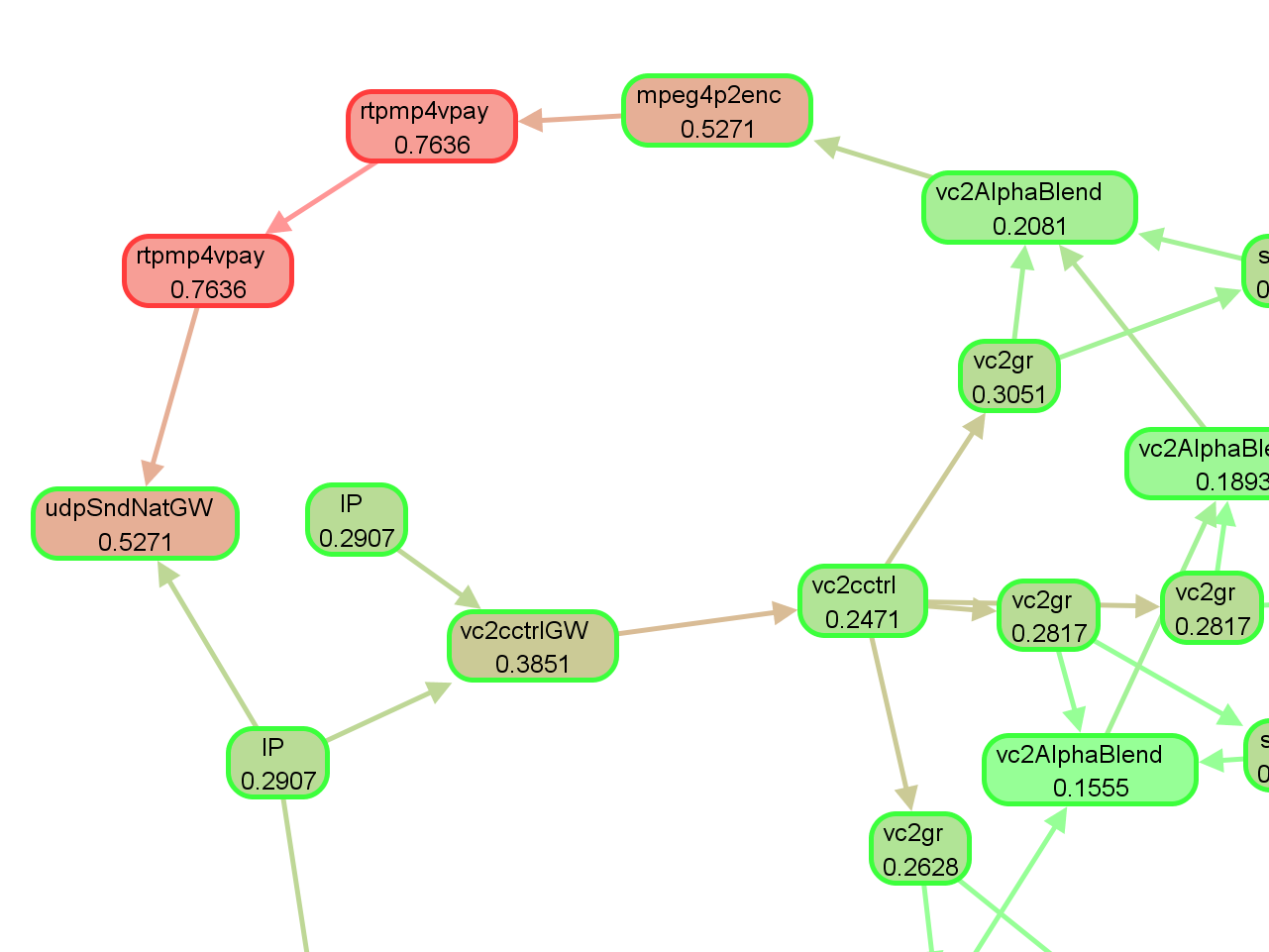}
 }

 \caption{Excerpts of visualizations of local predictions on graph structures, focused on anomalous substructures. The fill color of nodes and edges indicates the local threat prediction based on the classifier, while the borders of nodes indicate their, and intermediate edges' expected ground truth label. Shades of red correspond to an increased threat rating and correspond to the threat values displayed as floating point values on each node. In the left plot the anomaly is caused by the the node labeled \emph{crop} placed in between \emph{vc2gr} and \emph{scale}. The right plot shows an anomaly caused by the replication of the node labelled \emph{rtpmp4vpay}. }
 \label{fig:localpreds}
\end{figure}

For a SVDD classifier using a histogram intersection kernel we achieve perfect prediction accuracy, as visualized in the ROC-curves in in Figure \ref{fig:real_graphwise}. Even though the detection of anomalous graphs is flawless and without any false-positives neccesary, we can not gain perfect local predictions on individual graphs and edges. This is caused by the size of the substructures used during the feature extraction phase. In order to expressively capture the graph, the substructure features cover several connected nodes and edges, which limits the resolution of the local prediction approach introduced in Section \ref{sec:pred}. While larger neighborhood features might be able to even better describe the structure of a graph, the resolution for the local predictions might further diminish due to anomaly scores being projected on all covered nodes and edges, not only onto the covered part of the anomalous substructure, as visualized in Figure \ref{fig:localpreds}. Reducing the size of our neighborhood features would result in single nodes as substructures, which makes a classification of anomalous structures impossible. Predictions on the more complex synthetic data set seem less trivial, especially for the more complex anomaly types \emph{changed} and  \emph{mixed}. This slight loss in prediction accuracy, as well as the relative difficulty of the generated anomalies reflects well in the quality of the local predictions of both data sets. For those reason the 1st-degree neighborhood is an optimal solution.And although the detection performance for the synthetic data set, depicted in Figure \ref{fig:struct_rocs} is not optimal, those highlighted nodes are still perfectly usable by an analyzing administrator, since all of them are clustered around the real anomalous center nodes and can be interpreted as components potentially at risk.

Using the One-Class-SVM as a classifier yields comparable results for both data sets, as well as graph-wise and local predictions.

%% file: sections/evasion.tex
\subsection{Evasion}
\label{evasion}
When talking about anomaly detection methods, inevitably the risk of evasion needs to be addressed. The big assumption of anomaly detection methods is that the data, which is used to train the normal model, actually contains mostly normal data. Fortunately support vector machines are able to handle noisy data relatively robust and can be modified to be even more robust, as was shown by Biggio et.al. in \cite{BigNelLas11}. The most troublesome scenario occurs when using online learning, i.e. updating the model regularly, based on new data. Under those circumstances an adversary could iteratively poison the model, by increasing the amount of anomalous data in the normal data pool. But even for those most dangerous situations, an attack would fail due to impracticability, as was shown by Kloft et.al. in \cite{KloLas12}.

%% file: sections/conclusion.tex
\section{Conclusion}
\label{sec:conclusion}

In this paper we propose and evaluate two complimentary learning-based method for the detection of different dynamic and security-relevant anomalies in the Alcatel-Lucent Media Cloud. The proposed methods use behavioral and structural features and are also capable of locating the detected anomalies. In the evaluation we show on a data set of synthetic and real Media Cloud data perfect detection performances of up to $100\%$ true positive rate and very low false positive rates. This holds even when training on highly dynamic normal data. Furthermore we provide insight into the required learning and kernel methods, as well as additional advice on how to configure this system in different practical scenarios. Overall we conclude that our framework is a viable solution for detecting anomalies in Alcatel-Lucent Media Cloud Applications.

%

%% file: sections/acknowledgements.tex
\section{Acknowledgements}
\label{acknowledgements}
The authors would like to thank Ralf Klotsche for providing data and insights into the Alcatel-Lucent Media Cloud, as well as Konrad Rieck and Klaus-Robert M\"uller for fruitful discussions. The authors gratefully acknowledge funding from the German Federal Ministry of Education and Research (BMBF) under the project PROSEC (FKZ01BY1145).

%% file: master.bbl
\begin{thebibliography}{10}

\bibitem{CSA13.2}
C.~S. Alliance.
\newblock The notorious nine cloud computing top threats in 2013.
\newblock https://cloudsecurityalliance.org.

\bibitem{CSA13.1}
C.~S. Alliance.
\newblock Security guidance for critical areas of focus in cloud computing
  v3.0.
\newblock https://cloudsecurityalliance.org.

\bibitem{BarKliPloRon02}
P.~Barford, J.~Kline, D.~Plonka, and A.~Ron.
\newblock A signal analysis of network traffic anomalies.
\newblock In {\em Proceedings of the 2nd ACM SIGCOMM Workshop on Internet
  measurment}, pages 71--82. ACM, 2002.

\bibitem{BigNelLas11}
B.~Biggio, B.~Nelson, and P.~Laskov.
\newblock Support vector machines under adversarial label noise.
\newblock {\em Journal of Machine Learning Research-Proceedings Track},
  20:97--112, 2011.

\bibitem{ChaBanKum09}
V.~Chandola, A.~Banerjee, and V.~Kumar.
\newblock Anomaly detection: A survey.
\newblock {\em ACM Computing Surveys (CSUR)}, 41(3):15, 2009.

\bibitem{changrevisit}
W.-C. Chang, C.-P. Lee, and C.-J. Lin.
\newblock A revisit to support vector data description (svdd).

\bibitem{cortes1995support}
C.~Cortes and V.~Vapnik.
\newblock Support-vector networks.
\newblock {\em Machine learning}, 20(3):273--297, 1995.

\bibitem{CovKruVig10}
M.~Cova, C.~Kruegel, and G.~Vigna.
\newblock Detection and analysis of drive-by-download attacks and malicious
  javascript code.
\newblock In {\em Proceedings of the 19th international conference on World
  wide web}, pages 281--290. ACM, 2010.

\bibitem{CucObeSteDomMul13}
T.~Cucinotta, K.~Oberle, M.~Stein, P.~Domschitz, and S.~Mullender.
\newblock Run-time support for real-time multimedia in the cloud.
\newblock In {\em Proceedings of the 2nd International Workshop on Real-Time
  and Distributed Computing in Emerging Applications (REACTION 2013)}, 2013.

\bibitem{EbeHol07}
W.~Eberle and L.~Holder.
\newblock Anomaly detection in data represented as graphs.
\newblock {\em Intelligent Data Analysis}, 11(6):663--689, 2007.

\bibitem{eberle2007mining}
W.~Eberle and L.~B. Holder.
\newblock Mining for structural anomalies in graph-based data.
\newblock In {\em DMIN}, pages 376--389, 2007.

\bibitem{HidKas09}
S.~Hido and H.~Kashima.
\newblock A linear-time graph kernel.
\newblock In {\em Data Mining, 2009. ICDM'09. Ninth IEEE International
  Conference on}, pages 179--188. IEEE, 2009.

\bibitem{IngIno07}
K.~L. Ingham and H.~Inoue.
\newblock Comparing anomaly detection techniques for http.
\newblock In {\em Recent Advances in Intrusion Detection}, pages 42--62.
  Springer, 2007.

\bibitem{JanGra11}
W.~Jansen and T.~Grance.
\newblock Guidelines on security and privacy in public cloud computing.
\newblock https://cloudsecurityalliance.org.

\bibitem{KloLas12}
M.~Kloft and P.~Laskov.
\newblock Security analysis of online centroid anomaly detection.
\newblock {\em J. Mach. Learn. Res.}, 13(1):3681--3724, Dec. 2012.

\bibitem{LakCroDio05}
A.~Lakhina, M.~Crovella, and C.~Diot.
\newblock Mining anomalies using traffic feature distributions.
\newblock In {\em ACM SIGCOMM Computer Communication Review}, volume~35, pages
  217--228. ACM, 2005.

\bibitem{LiCroDioGovIanLak06}
X.~Li, F.~Bian, M.~Crovella, C.~Diot, R.~Govindan, G.~Iannaccone, and
  A.~Lakhina.
\newblock Detection and identification of network anomalies using sketch
  subspaces.
\newblock In {\em Proceedings of the 6th ACM SIGCOMM conference on Internet
  measurement}, pages 147--152. ACM, 2006.

\bibitem{LiuWanKumCha11}
W.~Liu, J.~Wang, S.~Kumar, and S.-F. Chang.
\newblock Hashing with graphs.
\newblock In {\em Proceedings of the 28th International Conference on Machine
  Learning (ICML-11)}, pages 1--8, 2011.

\bibitem{MatKumLat09}
T.~Mather, S.~Kumaraswamy, and S.~Latif.
\newblock {\em Cloud security and privacy: an enterprise perspective on risks
  and compliance}.
\newblock O'Reilly, 2009.

\bibitem{SchSmo02}
B.~Sch{\"o}lkopf and A.~Smola.
\newblock {\em Learning with Kernels}.
\newblock {MIT} Press, Cambridge, MA, 2002.

\bibitem{scholkopf1999support}
B.~Sch{\"o}lkopf, R.~C. Williamson, A.~J. Smola, J.~Shawe-Taylor, and J.~C.
  Platt.
\newblock Support vector method for novelty detection.
\newblock In {\em NIPS}, volume~12, pages 582--588, 1999.

\bibitem{SchBikKruRie12}
G.~Schwenk, A.~Bikadorov, T.~Krueger, and K.~Rieck.
\newblock Autonomous learning for detection of javascript attacks: vision or
  reality?
\newblock In {\em Proceedings of the 5th ACM workshop on Security and
  artificial intelligence}, pages 93--104. ACM, 2012.

\bibitem{ShaChr04}
J.~Shawe-Taylor and N.~Cristianini.
\newblock {\em Kernel methods for pattern analysis}.
\newblock Cambridge university press, 2004.

\bibitem{SulBonFurOrr13}
B.~Sullivan, E.~Bonver, J.~Furlong, and S.~Orrin.
\newblock Practices for secure development of cloud applications.
\newblock https://cloudsecurityalliance.org.

\bibitem{tax2004support}
D.~M. Tax and R.~P. Duin.
\newblock Support vector data description.
\newblock {\em Machine learning}, 54(1):45--66, 2004.

\bibitem{VelZla11}
D.~Velev and P.~Zlateva.
\newblock Cloud infrastructure security.
\newblock In {\em Open Research Problems in Network Security}, pages 140--148.
  Springer, 2011.

\bibitem{WanParSto06}
K.~Wang, J.~J. Parekh, and S.~J. Stolfo.
\newblock Anagram: A content anomaly detector resistant to mimicry attack.
\newblock In {\em Recent Advances in Intrusion Detection}, pages 226--248.
  Springer, 2006.

\bibitem{ZhaJueReiRis12}
Y.~Zhang, A.~Juels, M.~K. Reiter, and T.~Ristenpart.
\newblock Cross-vm side channels and their use to extract private keys.
\newblock In {\em Proceedings of the 2012 ACM conference on Computer and
  communications security}, pages 305--316. ACM, 2012.

\end{thebibliography}
